\documentclass[aip,
nofootinbib,
rsi,%
reprint,
twocolumn,
longbibliography,
author-numerical%
]{revtex4-2}

\usepackage[utf8]{inputenc}
\usepackage{booktabs}
\usepackage{siunitx}

\usepackage{amsfonts,amssymb,amsmath}
\usepackage[usenames]{color}
\usepackage{graphicx}
\usepackage{bm}
\usepackage{changes,comment}

\newcommand{\pdr}[2]{\dfrac{\partial{#1}}{\partial {#2}}}
\newcommand{\pddr}[2]{\dfrac{\partial^2{#1}}{\partial {#2}^2}}

\newcommand{\pdra}[2]{{\partial{#1}}/{\partial{#2}}}
\newcommand{\pddra}[2]{{\partial^2{#1}}/{\partial{#2}^2}}

\newcommand{\tx}{\tilde{x}}

\newcommand{\tZ}{\widetilde{Z}}
\newcommand{\tT}{\widetilde{T}}
\newcommand{\tR}{\widetilde{R}}

\newcommand{\tj}{\tilde{j}}
\newcommand{\teta}{\tilde{\eta}}

\newcommand{\Esig}{T_{\sigma}}

\newcommand{\tc}{\tilde{c}}
\newcommand{\cref}{c_{ref}}
\newcommand{\Cdl}{C_{dl}}

\newcommand{\veps}{\varepsilon}

\newcommand{\lexp}[1]{\exp\left(#1\right)}

\newcommand{\sion}{\sigma_p}

\newcommand{\lcat}{l_t}
\newcommand{\tit}{\tilde{t}}
\newcommand{\tom}{\tilde{\omega}}
\newcommand{\ri}{{\rm i}}
\newcommand{\expod}{{\rm e}^{\eta_0/b}}
\newcommand{\expo}{{\rm e}^{\teta_0}}

\newcommand{\Tleft}{T_{\rm CCL}}

\newcommand{\Tright}{T_{\rm FF}}

\newcommand{\etal}{{ }et al.{ }}

\begin{document}

\sf


\title{In-phase current and temperature oscillations reduce PEM fuel cell resistivity:\\ A modeling study}

\author{Andrei Kulikovsky}
\email{A.Kulikovsky@fz-juelich.de}

\affiliation{Forschungszentrum J\"ulich GmbH           \\
    Theory and Computation of Energy Materials (IET--3)   \\
    Institute of Energy and Climate Research,              \\
    D--52425 J\"ulich, Germany
}

\date{\today}

\begin{abstract}
We have developed a non-isothermal analytical model for the impedance of the cathode catalyst
layer (CCL) in a PEM fuel cell. In-phase harmonic perturbations to the current density
and temperature reduce the impedance and the static polarization resistivity of the CCL
due to lowering proton transport losses. A special selection of the current
and temperature perturbation amplitudes allows for complete elimination of these losses.
\end{abstract}


\maketitle

\section{Introduction}

Electrochemical impedance spectroscopy (EIS) of polymer electrolyte membrane (PEM)
fuel cells involves applying to the cell a small-amplitude harmonic signal, either
the current or potential, and measuring the resulting potential or current response.
EIS is one of the best and most popular methods of cell characterisation
(see Lasia's book~\cite{Lasia_book_14} and reviews~\cite{Zhang_20,Huang_20}).

However, much less work has been devoted to studying cell operating regime with
one of the operating parameters oscillating.
Experimental studies of Kim \etal\cite{Kim_08b} and Hwang \etal\cite{Hwang_10}
have demonstrated improved PEMFC performance under oscillating cathode flow velocity.
Modeling work of Kulikovsky \cite{Kulikovsky_24a,Kulikovsky_24d} have shown
that in-phase harmonic perturbations of the cell potential and oxygen concentration
lead to a dramatic reduction in PEMFC resistivity. In these works, the application
of the harmonic signal (or signals) aims to change the operating regime, rather
than measuring the impedance.


Below, we show that in-phase oscillations of the cell current density and temperature
decrease the impedance of the cathode catalyst layer (CCL).
A pair of ``potential--oxygen concentration''
perturbations lowers the CCL resistivity by decreasing the oxygen transport
losses in the CCL  \cite{Kulikovsky_24a,Kulikovsky_24d}.
Similarly, the pair of ``current--temperature'' perturbations
reduces the CCL resistivity by lowering the proton transport losses
due to the Arrhenius dependence of proton conductivity on temperature.

\section{Model}

\begin{figure}
\begin{center}
   \includegraphics[scale=1.1]{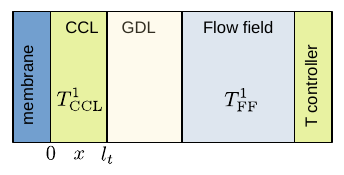}
\end{center}
\caption{Schematic of the cell cathode side. CCL, GDL and FF stand for the cathode catalyst layer,
     gas diffusion layer, and the flow field, respectively. $\Tleft^1$ and $\Tright^1$ indicate
     the AC perturbation amplitudes of the CCL and FF temperatures.
}
\label{fig:one}
\end{figure}

A schematic of the cell cathode side is shown in Figure~\ref{fig:one}.
 For brevity, the application
of temperature oscillations will be referred to as temperature control.
A model of heat transport through the cell cathode side leads to linear
relations between small perturbations of the CCL and flow field (FF)
temperatures, namely $\Tleft^1$, $\Tright^1$ (to be published elsewhere).
The external surface
of the FF can be used for temperature control.
These linear relations mean that the applied small
harmonic perturbation $\Tright^1$ causes a small harmonic perturbation
of the CCL temperature $\Tleft^1$.
The CCL proton conductivity $\sion$ depends on the temperature via the Arrhenius law (see below).
Thus, by modulating the CCL temperature, we modulate the proton conductivity.

To study the effect of $\sion$ oscillations on CCL impedance, we will assume
that oxygen transport loss in the CCL is negligible.
This assumption is valid if the cell current is sufficiently small.
In that case, the proton charge conservation equation is
\begin{equation}
   \Cdl\pdr{\eta}{t} - \sion\pddr{\eta}{x}
       = - i_*\left(\dfrac{c_1}{\cref}\right)\lexp{\dfrac{\eta}{b}}
   \label{eq:etax}
\end{equation}
Here,
$\Cdl$ is the double layer capacitance,
$\eta$ the ORR overpotential, positive by convention,
$\sion$ the CCL proton conductivity,
$i_*$ the ORR volumetric exchange current density,
$c_1$ the oxygen molar concentration in the CCL,
which is assumed to be uniform,
$\cref$ the reference oxygen concentration, and
$b$ the ORR Tafel slope.
For further references we note that the static equation for $\eta^0$ is
\begin{equation}
   \sion^0\pddr{\eta^0}{x} =  i_*\left(\dfrac{c_1}{\cref}\right)\expod.
   \label{eq:eta0x}
\end{equation}
Here and below, the subscripts 0 and 1 mark the static variables and the small perturbation
amplitudes, respectively.

Following Springer \etal \cite{Springer_91}, the temperature dependence $\sion(T)$ is given
by the Arrhenius law
\begin{equation}
   \sigma_p(T) = \sion^*\exp\left(\Esig\left(\dfrac{1}{303} - \dfrac{1}{T}\right) \right),
      \quad \Esig = 1268~\text{K.}
   \label{eq:sionT}
\end{equation}
Setting $\sion = \sion^0 + \sion^1$,
$T = T^0 + T^1$, expanding the exponent in a Taylor series,
and retaining the two leading terms, we get the relationship
between the temperature $T^1$ and the proton conductivity $\sion^1$ perturbations:
\begin{equation}
   \sion^1 = \sion^0\dfrac{\Esig T^1}{\left(T^0\right)^2},
      \quad \sion^0 \equiv \sigma_p(T^0)
   \label{eq:sion1}
\end{equation}
Here, $T^0$ is the steady-state cell temperature, which is assumed to
be uniform across the cathode side. Small non-uniformity in $T^0$
does not play any role in the present analysis.
Evidently, due to the linearity of Eq.\eqref{eq:sion1} in $T^1$, it holds
for the perturbation amplitudes $\sion^1(\omega)$ and $T^1(\omega)$ in the frequency domain.

Setting in Eq.\eqref{eq:etax} $\eta = \eta^0(x) +\eta^1(x,t)$, $\sion = \sion^0 + \sion^1$,
expanding exponent in a Taylor series, neglecting the term with the perturbation products,
and subtracting the static equation for $\eta^0$, Eq.\eqref{eq:eta0x}, we get
\begin{equation}
   \Cdl\pdr{\eta^1}{t} - \sion^1\pddr{\eta^0}{x} - \sion^0\pddr{\eta^1}{x}
       = - i_*\left(\dfrac{c_1}{\cref}\right)\expod\dfrac{\eta^1}{b}
   \label{eq:eta1x}
\end{equation}
Substituting here the Fourier-transform \\ $\eta^1(x,t) = \eta^1(x, \omega)\exp(\ri\omega t)$,
we obtain
the equation for the perturbation amplitude $\eta^1(x,\omega)$
\begin{equation}
   \Cdl\ri\omega\eta^1 - \sion^1\pddr{\eta^0}{x} - \sion^0\pddr{\eta^1}{x}
       = - i_*\left(\dfrac{c_1}{\cref}\right)\expod\dfrac{\eta^1}{b}
   \label{eq:eta1xF}
\end{equation}
Using Eq.\eqref{eq:eta0x} to eliminate the second derivative $\pddra{\eta^0}{x}$,
substituting Eq.\eqref{eq:sion1} and rearranging the terms, we find
\begin{equation}
   \sion^0\pddr{\eta^1}{x}  =  \Cdl\ri\omega\eta^1
       + \left(\dfrac{\eta^1}{b} - \dfrac{\Esig T^1}{\left(T^0\right)^2}\right)
                                    i_*\left(\dfrac{c_1}{\cref}\right)\expod
   \label{eq:eta1xF3}
\end{equation}
The boundary conditions for Eq.\eqref{eq:eta1xF3} are
\begin{equation}
   \left(- \sion^0 \pdr{\eta^1}{x} - \sion^1 \pdr{\eta^0}{x}\right)_{x=0}  = j^1, \quad
   \left.\pdr{\eta^1}{x}\right|_{x=\lcat} = 0
   \label{eq:bc1}
\end{equation}
where
and the first equation follows from linearisation of the Ohm's law
$(\sion^0 +\sion^1)\pdra{(\teta^0 + \teta^1)}{\tx} = - j_1$ and $\lcat$ is the CCL thickness.
Taking into account the static Ohm's law $- \sion^0\pdra{\eta^0}{x} = j^0$ and Eq.\eqref{eq:sion1}, equation~\eqref{eq:bc1} is transformed to
\begin{equation}
   - \left.\sion^0 \pdr{\eta^1}{x} \right|_{x=0}  = j^1 - \dfrac{\Esig T^1}{\left(T^0\right)^2}j^0, \quad
   \left.\pdr{\eta^1}{x}\right|_{x=\lcat} = 0
   \label{eq:bc3}
\end{equation}

To simplify further calculations, we introduce the dimensionless variables
\begin{multline}
   \tx = \dfrac{x}{\lcat}, \quad \tit = \dfrac{t}{t_*}, \quad \teta = \dfrac{\eta}{b},
   \quad \tj = \dfrac{j}{j_*}, \quad \tc = \dfrac{c}{\cref}, \\
   \tT = \dfrac{T}{\Esig}, \quad \tom = \omega t_*, \quad \tZ = \dfrac{Z \sion}{\lcat}
   \label{eq:dless}
\end{multline}
where
$\omega$ the AC angular frequency,
$Z$ the impedance, and
\begin{equation}
    t_* = \dfrac{\Cdl b}{i_*}, \quad j_* = \dfrac{\sion^0 b}{\lcat}.
    \label{eq:tast}
\end{equation}
are the characteristic time and current density, respectively.

With these variables, Eqs.\eqref{eq:eta1xF3}, \eqref{eq:bc1} take the form
\begin{equation}
   \veps^2\pddr{\teta^1}{\tx}  =  \ri\tom\teta^1
       + \left(\teta^1 - \Lambda^1\right) \tc_1\expo
   \label{eq:teta1xF3}
\end{equation}
\begin{equation}
   - \left. \pdr{\teta^1}{\tx} \right|_{\tx=0}  = \tj^1 - \Lambda^1\tj^0, \quad
   \left.\pdr{\teta^1}{\tx}\right|_{\tx=1} = 0
   \label{eq:tbc3}
\end{equation}
where
\begin{equation}
   \Lambda^1 = \tT^1 \bigg/ \left(\tT^0\right)^2, \quad  \veps = \sqrt{\dfrac{\sion^0 b}{i_* \lcat^2}}
   \label{eq:Lam1}
\end{equation}

\section{Results and discussion}

Let the cell current density be small. In that case, we can neglect the variation of the ORR
overpotential through the CCL depth and set $\teta^0 = \teta_0$, where
$\teta_0$ is the overpotential at the membrane surface (at $\tx=0$).
The subscripts 0 and 1 denote the membrane/CCL and CCL/GDL interfaces, respectively.
The dimensionless Tafel law
\begin{equation}
   \veps^2 \tj_0 = \tc_1\expo
   \label{eq:tTafel}
\end{equation}
allows us to replace  $\tc_1\expo$  in Eq.\eqref{eq:teta1xF3} with $\veps^2 \tj_0$.
The solution to Eq.\eqref{eq:teta1xF3} with the boundary conditions Eq.\eqref{eq:tbc3} is
\begin{multline}
   \teta^1(\tx) = \dfrac{\Lambda^1\tj^0 - \tj^1}{\phi}\bigl(\sinh(\phi\tx) - \coth(\phi) \cosh(\phi\tx)\bigr) \\
               + \dfrac{\tj_0 \Lambda^1}{\phi^2}, \quad \phi = \sqrt{\tj_0 + \ri\tom/\veps^2}
   \label{eq:teta1_sol}
\end{multline}
Eq.\eqref{eq:teta1_sol} gives the system impedance $\tZ = \teta^1/\tj^1 |_{\tx=0}$:
\begin{equation}
   \tZ = (1 - \kappa)\dfrac{\coth\sqrt{\tj_0 + \ri\tom/\veps^2}}{\sqrt{\tj_0 + \ri\tom/\veps^2}}
   + \dfrac{\kappa}{\tj_0 + \ri\tom/\veps^2}
   \label{eq:tZtot}
\end{equation}
where $\kappa$ is the temperature control parameter:
\begin{equation}
   \kappa =  \dfrac{\Lambda^1\tj_0}{\tj^1}
   \label{eq:kap}
\end{equation}
Since we can choose the values of  $\tj^1$ and $\Lambda^1$,
the control parameter $\kappa$ can always be taken
in the range $\kappa \in [0, 2]$ studied in this work.

\begin{figure}
\begin{center}
   \includegraphics[scale=0.45]{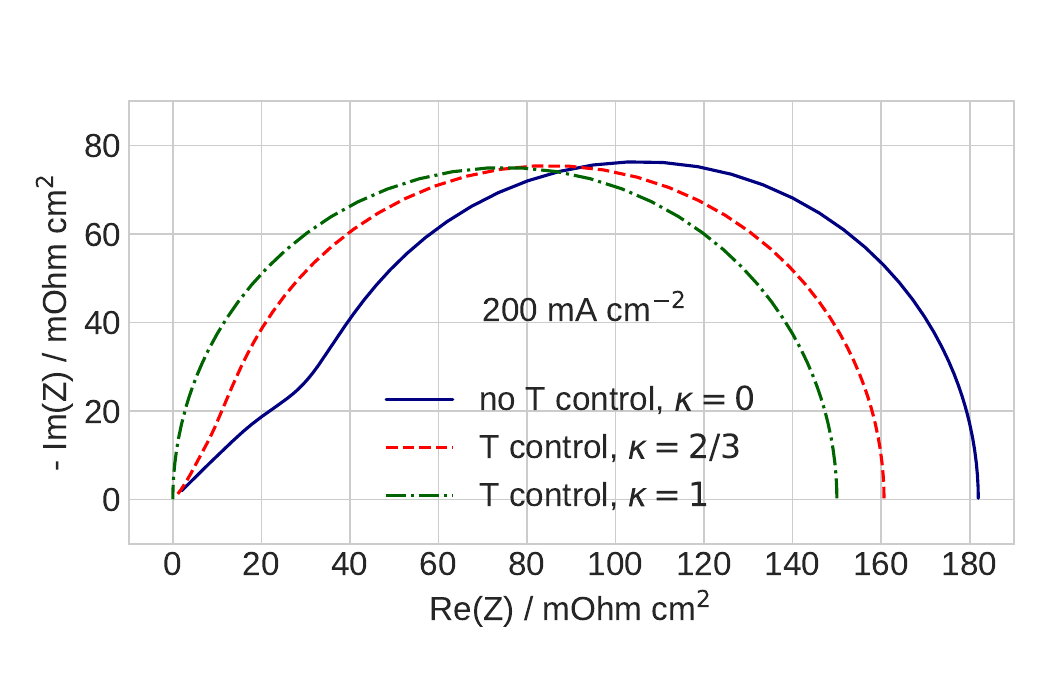}
\end{center}
\caption{The Nyquist spectra of impedance Eq.\eqref{eq:tZtot} for three values
     of the temperature control parameter $\kappa$. The cell current density
     is 200 mA~cm$^{-2}$, the other parameters are listed in Table~\ref{tab:parms}.
}
\label{fig:Nyq200}
\end{figure}

At zero temperature perturbation, $\Lambda^1=0$, $\kappa=0$,
and Eq.\eqref{eq:tZtot} reduces to the impedance of the CCL with
the finite rate of proton transport \cite{KulikovskY_13a}:
\begin{equation}
   \tZ_{fp} = \dfrac{\coth\sqrt{\tj_0 + \ri\tom/\veps^2}}{\sqrt{\tj_0 + \ri\tom/\veps^2}}
   \label{eq:tZccl}
\end{equation}
The typical shape of the Nyquist spectrum of $Z_{fp}$ for the set of parameters
in Table~\ref{tab:parms} is shown in Figure~\ref{fig:Nyq200} (the blue solid curve).
It comprises the faradaic arc connected to the high-frequency,  $45^\circ$ straight line,
which exhibits the proton transport impedance.

\begin{table}
\small
\begin{tabular}{|l|c|}
\hline
     ORR exchange current density $i_*$,  A~cm$^{-3}$ &  $ 10^{-4}$ \\
     Double layer capacitance $\Cdl$, F~cm$^{-3}$ &  20 \\
     ORR Tafel slope $b$, mV/exp               & 30 \\
     CCL proton conductivity $\sion^0$, $\Omega^{-1}$~cm$^{-1}$  & 0.01 \\
     CCL thickness $\lcat$, $\mu$m           & 10   \\
\hline
   	 Cathode pressure $p$, bar                 & 1.0 \\
     Cell temperature $T^0$, K                 & $273 + 80$ \\
     Cell current density $j_0$, A~cm$^{-2}$   &  0.2, 0.413  \\

\hline
\end{tabular}
\caption{The cell properties and operation parameters.}
\label{tab:parms}
\end{table}

Now consider the case of non-zero temperature perturbation, $\kappa > 0$.
For a moment, we leave aside the question of if and how the high-frequency temperature
perturbation could be produced.

The Nyquist plots of the impedance $Z$, Eq.\eqref{eq:tZtot},
for $\kappa$ of 2/3 and 1 are shown in Figure~\ref{fig:Nyq200}a. As can be seen,
the positive $\kappa$ lowers the static cell resistivity.
This effect is best seen with $\kappa = 1$. In that case, Eq.\eqref{eq:tZtot} reduces to
\begin{equation}
   \tZ_{RC} = \dfrac{1}{\tj_0 + \ri\tom/\veps^2},
   \label{eq:tZrc}
\end{equation}
which is the parallel $RC$-circuit impedance. The static resistivity $\tR_{RC}$
of Eq.\eqref{eq:tZrc} is independent of the proton conductivity:
\begin{equation}
   R_{RC} = \dfrac{b}{j_0}
   \label{eq:Rrc}
\end{equation}
With the parameters in Table~\ref{tab:parms}, $R_{RC} = 150$ mOhm~cm$^{2}$, which
is noticeably smaller than 180  mOhm~cm$^{2}$ at $\kappa=0$
(Figure~\ref{fig:Nyq200}a). Therefore, temperature oscillations in phase with
the oscillations of the cell current density can reduce or even
eliminate the proton transport losses in the CCL.

This effect is analogous to parametric resonance in mechanical systems.
The analysis above is performed under the assumption that $\kappa$ is real and positive,
i.e.,  $\tT^1$ is in phase with $\tj^1$, Eq.\eqref{eq:kap}.
According to Eq.\eqref{eq:sion1}, the resulting proton conductivity oscillations
are also in phase with the $j^1$ oscillations. Writing Ohm's law in the form
\begin{equation}
   - \sion^0 \pdr{\eta^1}{x} = j^1 + \sion^1 \pdr{\eta^0}{x}
                             = j^1 - \dfrac{\sion^1}{\sion^0} j_0
   \label{eq:Ohms}
\end{equation}
we see that the conductivity oscillations $\sion^1$ reduce
the amplitude of $\pdra{\eta^1}{x}$ oscillations. Furthermore,
when $\sion^1 j_0/\sion^0 = j^1$, we get $\pdra{\eta^1}{x} = 0$.
A uniform along $\tx$ overpotential perturbation $\teta^1$
means zero proton transport losses in the CCL.
The in-phase oscillations of $\sion^1$ and $\tj^1$
justify the analogy with parametric resonance.

At higher cell currents, the effect can be studied using Eq.\eqref{eq:teta1xF3}
with the non-uniform along $\tx$ static overpotential $\teta^0(\tx)$. The latter
is obtained by solving Eq.\eqref{eq:eta0x}, which, in the dimensionless form, reads
\begin{equation}
   \veps^2\pddr{\teta^0}{\tx} =  \tc_1\expo,
      \quad \teta^0(0) = \teta_0, \quad \left.\pdr{\teta^0}{\tx}\right|_{\tx=0} = 0
   \label{eq:teta0x}
\end{equation}
The solution of the BVP problem Eq.\eqref{eq:teta0x}
should be substituted into Eq.\eqref{eq:teta1xF3}. The numerical
solution to Eq.\eqref{eq:teta1xF3} yields
the system impedance $\tZ = \teta^1/\tj^1 |_{\tx=0}$ at higher currents.
The details of this procedure will be reported in a full-length paper.

\begin{figure}
\begin{center}
   \includegraphics[scale=0.45]{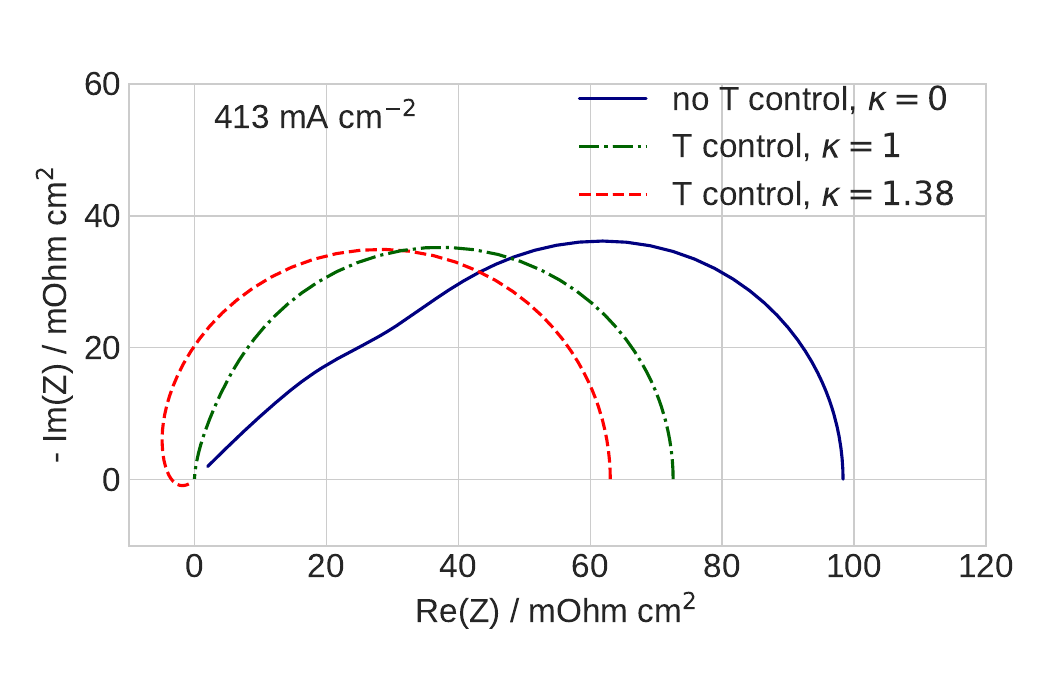}
\end{center}
\caption{The Nyquist spectra of CCL impedance resulting from the numerical solution of Eq.\eqref{eq:teta1xF3}
     with the $\tx$--dependent static overpotential $\teta^0(\tx)$
     for three values of the temperature control parameter $\kappa$. The cell current density
     is 413 mA~cm$^{-2}$.
}
\label{fig:Nyq413}
\end{figure}

Figure~\ref{fig:Nyq413} shows the Nyquist spectra for a cell current density of 0.413 A~cm$^{-2}$,
obtained from the numerical solution of Eq.\eqref{eq:teta1xF3} with the boundary conditions
Eq.\eqref{eq:tbc3} and $\teta^0(\tx)$ resulting from Eq.\eqref{eq:teta0x}. Without
the temperature control ($\kappa=0$), the spectrum resembles a Warburg finite-length
impedance, similar to that in Figure~\ref{fig:Nyq200}
(cf. the solid blue curves in Figures~\ref{fig:Nyq413} and \ref{fig:Nyq200}). With the control
parameter $\kappa=1$, the spectrum transforms into an almost ideal semicircle,
without the proton transport straight line in the high-frequency domain
(the dash-dotted green curve in Figure~\ref{fig:Nyq413})). When the control
parameter exceeds unity, $\kappa = 1.37$, the spectrum further shrinks
and a new feature arises: an inductive-like
high frequency loop (the dashed red curve in Figure~\ref{fig:Nyq413}).

%

The most striking feature of the curves in Figures~\ref{fig:Nyq200} and \ref{fig:Nyq413} is that
the {\em static} cell resistivity (the rightmost
point of the Nyquist spectrum) decreases as $k$ increases. This would perhaps be less surprising if we note
that the effect of temperature control does not vanish at a vanishingly small $\omega$,
which, from the practical perspective, does not differ much from $\omega=0$.

Due to the inertia of heat transport through the FF and GDL,
it is hardly feasible to produce high-frequency $\Tleft^1 \equiv T^1$ oscillations
in experiment. However, Figures~\ref{fig:Nyq200} and \ref{fig:Nyq413} show
a reduction in cell resistivity at low frequencies. The low-frequency $T^1$
perturbation can be arranged  by means of a temperature controller mounted
on the external side of the FF (Figure~\ref{fig:one}). The AC signal of the controller should
account for the phase shift due to heat transport through the FF and GDL.
The experimental study of the effect can be performed in the frequency range below 0.1 Hz.

The reduction of the cell resistivity upon simultaneous application of potential
and oxygen concentration perturbations to the PEM fuel cell
has been demonstrated in modeling works \cite{Kulikovsky_24a,Kulikovsky_24d}.
Here, we see that the proper temperature control eliminates the proton transport losses
in the CCL. In both cases, operating the cell in an EIS-like regime
is promising from a practical perspective.



\end{document}